\documentclass[conference]{IEEEtran}
\IEEEoverridecommandlockouts
\usepackage{cite}
\usepackage{balance}
\usepackage{amsmath,amssymb,amsfonts}
\usepackage{algorithm2e}
\usepackage{graphicx}
\usepackage{textcomp}
\usepackage{xcolor}
\def\BibTeX{{\rm B\kern-.05em{\sc i\kern-.025em b}\kern-.08em
    T\kern-.1667em\lower.7ex\hbox{E}\kern-.125emX}}

\usepackage{cleveref}
\DeclareMathOperator*{\argmax}{arg\,max}
\DeclareMathOperator*{\argmin}{arg\,min}
\usepackage[bottom]{footmisc}

\usepackage{url}
\usepackage{comment}
\usepackage{makecell}
\usepackage{float} 
\Crefname{figure}{Fig.}{Figs.}
\Crefname{section}{Sec.}{Secs.}
\usepackage{subcaption}

\begin{document}
\bstctlcite{IEEEexample:BSTcontrol} 

\title{Extended FastSLAM Using Cellular Multipath Component Delays and Angular Information}
\author{
\IEEEauthorblockN{Junshi Chen\IEEEauthorrefmark{2}\IEEEauthorrefmark{1}, Russ Whiton\IEEEauthorrefmark{2}\IEEEauthorrefmark{3}, Fredrik Tufvesson\IEEEauthorrefmark{2}}
\IEEEauthorblockA{\IEEEauthorrefmark{2}Dept. of Electrical and Information Technology, Lund University, Lund, Sweden\\
\IEEEauthorblockA{\IEEEauthorrefmark{1}Terranet AB, Lund, Sweden}
\IEEEauthorblockA{\IEEEauthorrefmark{3}Volvo Car Corporation, SE-405 31 Gothenburg, Sweden}
Email: \{junshi.chen, russell.whiton, fredrik.tufvesson\}@eit.lth.se}
}
\cleardoublepage
\maketitle
\begin{abstract}
 %Positioning using wireless signals is attractive in many scenarios when other sensors are facing challengings. In this paper, the long-term evolution (LTE) downlink signals from two neighboring commercial base stations (BS) in Lund city of Sweden are received by a massive antenna array mounted on a passenger vehicle. Multipath component (MPC) delays and angular-of-arrival (AOA) information extracted from the received signals are used to estimate the positions of the vehicle, the transmitters, and the virtual transmitters (VT) with the extended fast simultaneous localization and mapping (FastSLAM) algorithm. The results show that the algorithm can accurately estimate the positions of the vehicle and the transmitters (and virtual transmitters) accurately. The vehicle's positioning error is less than 6 meters in a traversed  distance of 530 meters.
 Opportunistic navigation using cellular signals is appealing for scenarios where other navigation technologies face challenges. In this paper, long-term evolution (LTE) downlink signals from two neighboring commercial base stations (BS) are received by a massive antenna array mounted on a passenger vehicle. Multipath component (MPC) delays and angle-of-arrival (AOA) extracted from the received signals are used to jointly estimate the positions of the vehicle, transmitters, and virtual transmitters (VT) with an extended fast simultaneous localization and mapping (FastSLAM) algorithm. The results show that the algorithm can accurately estimate the positions of the vehicle and the transmitters (and virtual transmitters). The vehicle's horizontal position error of SLAM fused with proprioception is less than 6 meters after a traversed distance of 530 meters, whereas un-aided proprioception results in a horizontal error of 15 meters.
\end{abstract}

% Russ' attempt at Introduction
\section{Introduction}
% Russ' attempt at introduction using flow of paragraphs (ITS -> Cellular -> SLAM)
Intelligent transportation systems hold promise for traffic safety and efficiency. Localization performance is important for such systems, from use cases ranging from traffic optimization \cite{cvetek2021survey} to autonomous driving \cite{Reid2019LocalizationRF}. A broad sensor suite has been developed for the localization problem for these challenging use cases, but the limitations like weak signal strength and multipath in the city urban for the global navigation system (GNSS), and the dim light for the camera and lidar systems, etc., have motivated efforts toward finding additional sensors to provide localization information.

Cellular communication also has a long history of use for localization as an alternative or complement to satellite-based navigation \cite{del20181to5gpos}, and cellular technologies are expected to further develop towards joint communication and sensing \cite{henk2021icas}. Not only does this offer benefits for optimizing communication system performance \cite{Koivisto2017locaawarecomm}, but also can play a role in addressing even the most demanding localization use cases such as autonomous driving \cite{bagher5gv2x}. The manner in which the cellular signals are generated and utilized for positioning can take many forms, and various classification schemes have been suggested, e.g. in \cite{whiton2022cellular}, it is categorized into three types: triangulation and multilateration, machine learning based positioning, and simultaneous localization and mapping (SLAM) \cite{Durrant-Whyte2006ram}.

SLAM has multiple uses, including for tracking purposes \cite{xuhong2019mmimopos} or in augmenting proprioception sensors on vehicles \cite{whiton2022cellular}, but successful implementation requires associating measurements from different snapshots, which can be difficult depending on the type of sensor used and data quality. Significant effort has been spent on achieving accurate data association, and many advanced algorithms have been developed, e.g., joint probability data association (JPDA) \cite{Fortmann1983jpda}, and belief propagation (BP) \cite{erik_bp_ieee}. On the other hand, FastSLAM \cite{thrun2005pr} takes another approach to simplify the data association problem. It uses a particle filter mechanism, and for each particle, only a simple maximum likelihood data association is applied independently. During the update of the FastSLAM, only the particles with the highest likelihood of data association can survive. In this way, it can keep the effective data association in general.

In this paper, the FastSLAM algorithm is applied to the parameters extracted from the cellular signals including multipath component (MPC) delays, azimuthal angle-of-arrival (AOA), and elevational AOA. The parameters are extracted from the signals of multiple commercial long-term evolution (LTE) base stations (BS) received by a massive antenna array mounted on a passenger vehicle in an urban environment. FastSLAM is extended to work with observations from multiple antenna ports and BSs. This extended FastSLAM can further simplify the data association problem by further factoring the posterior model and processing each port independently. Results from field measurements show that the extended FastSLAM algorithm works well in complicated urban environments, and the vehicle’s positioning error is less than 6 meters after a traversed distance of 530 meters.

The structure of the paper is as follows. \Cref{sec:system_model} introduces the wireless signal system model. \Cref{sec:fastslam_model} describes the extended FastSLAM model using the estimated MPC parameters, \Cref{sec:fastslam_update} describes the iterative update of the extended FastSLAM algorithm applied to localize the positions of the vehicle, the transmitters and the virtual transmitters. \Cref{sec:measure} presents the measurement setup and analysis of the results from the measurement data and the proposed algorithm. Finally, \Cref{sec:summary} summarizes the paper. 

\textit{Notation}: Matrices and vectors are denoted as uppercase and lowercase boldface letters, e.g., $\mathbf{A}$ and $\mathbf{a}$. The identity matrix is denoted as $\mathbf{I}$. The matrix transpose and matrix inverse are denoted as superscripts $ (\cdot)^T$ and $(\cdot)^{-1}$ respectively. The Euclidian norm is denoted as $\left\Vert \cdot \right\Vert$. The speed of light is $ c \simeq 3 \cdot 10^8$ m/s. 
\section{system model}\label{sec:system_model}
In LTE systems, orthogonal frequency division multiplexing (OFDM) is used and the baseband signal transmitted from one antenna port of one BS is described as \cite{junshi2022ic_rimax}
\begin{equation}
\begin{split}
s^{j,k}(t) &= \sum_{n=-N_{sc}/2}^{n=-1}{x}^{j,k}[n+N_{sc}/2]e^{i2\pi n \Delta ft}\\
&+\sum_{n=0}^{n=N_{sc}/2-1}{x}^{j,k}[n+N_{sc}/2]e^{i2\pi (n+1) \Delta ft}
\end{split}
\end{equation}
here ${x}^{j,k}[n],\, n\in \{ 0,N_\text{sc}-1 \}$ is the transmitted signal at the $n$-th subcarrier from the $ j$-th antenna port of the $k$-th BS, $ j \in \{ 1,\ldots,4\} $ is the antenna port number of the cell-specific reference symbol (CRS), $k\in \{1,\ldots,K\} $ is the number of BSs transmitting signals, and $N_{sc}$ is the number of subcarriers in the OFDM symbol. Further, $t$ is limited to $[-T_\text{CP},T_\text{s}]$ denoting continuous time, $T_\text{CP}$ is the duration of the cyclic prefix (CP), and $ T_\text{s} = 1/\Delta f $ is
the duration of one OFDM symbol with $ \Delta f$ being the subcarrier spacing. CRS ${x}_{CRS}^{j,k}$ are transmitted on specific subcarriers and symbols depending on the cell ID, antenna port number, CP type, and bandwidth of the LTE system \cite{3gpp36211}. CRS are appealing for positioning because, unlike synchronization signals, they span the full channel bandwidth (giving time resolution) and are transmitted more frequently. In this paper, they are used exclusively to estimate the position of the vehicle and those of the transmitters and virtual transmitters. %In this paper, the focus is on using the extracted parameters from the CRS symbols to estimate the position of the vehicle and those of the transmitters and virtual transmitters, therefore, other symbols and channels are ignored.

A 128-port stacked uniform circular antenna array is used with the receiver. The antennas are switched in a fixed sequence with a switching interval of 0.5 ms, and all 128 ports are sampled for a complete snapshot every 75 ms including 11 ms for automatic gain control. The receiver moves at a relatively low average speed of 1.0 m/s because of the switched nature of the measurement system \cite{junshi2022ic_rimax}. 

The channel frequency response from the $j$-th port of the $k$-th BS is modeled as a summation of $M$ MPCs parameterized by their delay $\tau^{m,j,k}$, direction-of-arrival (DOA) $ \Omega^{m,j,k} $, and Doppler shift $\nu^{m,j,k} $. The DOA is further divided into azimuth AOA $\varphi^{m,j,k}$ and elevation AOA $\theta^{m,j,k}$. The time-varying directional transfer function at the $n$-th subcarrier is represented as
\begin{equation}
\resizebox{.91\hsize}{!}{$
\mathbf{h}^{j,k}[n] = \sum_{m=1}^M\mathbf{b}_R({\Omega}^{m,j,k}) {\boldsymbol{\Gamma}}^{m,j,k}\mathbf{b}_{T}^{j,k}e^{-i2\pi(n\Delta f\tau^{m,j,k}-\nu^{m,j,k}t)}\\
$}
\end{equation}
where $\mathbf{b}_R({\Omega}^{m,j,k})\in \mathbb{C}^{128 \times 2}$ is the receive antenna array pattern, $\mathbf{b}_{T}^{j,k}\in \mathbb{C}^{2 \times 1} $ is the $j$-th port of the $k$-th BS antenna response. ${\boldsymbol{\Gamma}}^{m,j,k}$ is the polarimetric path weight matrix defined as
\begin{equation}
    {\boldsymbol{\Gamma}}^{m,j,k} = \begin{bmatrix}
{\gamma}_{\text{HH}}^{m,j,k} & {\gamma}_{\text{VH}}^{m,j,k} \\ 
{\gamma}_{\text{HV}}^{m,j,k} & {\gamma}_{\text{VV}}^{m,j,k} 
\end{bmatrix}.
\end{equation}
The matrix elements represent different polarization combinations of the transmitter and the receiver, e.g., $\text{HV}$ is horizontal-to-vertical. 

The aggregate received CRS in the frequency domain at the $n$-th subcarrier is given as follows 
\begin{equation}
\mathbf{y}[n]=\sum_{k=1}^K\sum_{j=1}^J\mathbf{h}^{j,k} [n] \cdot x_{CRS}^{j,k}\left[ n\right].
\end{equation}
Using the channel parameter estimation and interference cancellation methods described in \cite{junshi2022ic_rimax}, signals from different antenna ports of different BSs are separated and the MPC parameters delay, azimuth AOA, elevation AOA and signal-to-noise ratio (SNR) are estimated by the improved RIMAX algorithm. 

The estimated MPC parameters can be used for positioning, and for this purpose, the MPC delays are converted into the distance domain by adding the unknown and fixed clock offset between the $k$-th BS and the vehicle $t^k_{\textrm{offset}}$ and then multiplying with the speed of light  
\begin{equation}
    d^{m,j,k}(t) = \left(\tau^{m,j,k}(t) +t^k_{\textrm{offset}}\right)\cdot c.
\end{equation}

The estimated parameters of all the MPCs at time index $t$ can be represented as
\begin{align}
\begin{split}
    \mathbf{Z}_t =& \left[\mathbf{z}^{1,1,1}(t),\ldots, \mathbf{z}^{M,J,K}(t)\right] 
\end{split}\\
\begin{split}
    \mathbf{z}^{m,j,k}(t)=& \left[ 
 d^{m,j,k}(t), \varphi^{m,j,k}(t),
 \theta^{m,j,k}(t) \right]^T
\end{split}
\end{align}
and the SNR of all MPCs at time index $t$ can be represented as
\begin{equation}
\boldsymbol{\lambda}_t = \left[\lambda^{1,1,1}(t),\ldots , \lambda^{M,J,K}(t)\right].
\end{equation}
\section{Exended FastSLAM model using the estimated MPC parameters}\label{sec:fastslam_model}
MPCs from BSs with direct line-of-sight (LOS), or MPCs from reflectors and scatters in the environment with non line-of-sight (NLOS) are considered as synchronized and independent transmitters and virtual transmitters (VT) respectively \cite{Gentner2016twc}. For convenience of representation, the term VT is used to refer to all transmitters. The problem to be solved is to use all the parameters extracted from the wireless signals and fuse with the velocity information from the vehicle to estimate the positions of the vehicle and the VTs accurately, and also associate VTs across measurements at different times. The posterior can be represented as 
\begin{equation}
p\left(\mathbf{V},\mathbf{c}_{1:t}, \mathbf{r}_{1:t}\mid\mathbf{Z}_{1:t},\mathbf{u}_{1:t}\right)   
\end{equation}
here $\mathbf{V}$ represents positions of the VTs, $\mathbf{r}_{1:t}$ is the time series of the vehicle state vector, $\mathbf{Z}_{1:t}$ are the measurements, $\mathbf{c}_{1:t}$ is the association between measurements and VTs, and $\mathbf{u}_{1:t}$ is the input velocity from the vehicle. The index $1:t$ represents the time from time index $1$ to $t$.

The positions of the VTs are given as 
\begin{align}
\begin{split}
\mathbf{V} &= \left[\mathbf{v}^1,\ldots, \mathbf{v}^{L,J,K} \right]
\end{split}\\
\begin{split}
    \mathbf{v}^{l,j,k} &= \left[v_x^{l,j,k}, v_y^{l,j,k}, v_z^{l,j,k}\right]^T
\end{split}
\end{align}
here $\mathbf{v}^{l,j,k}$ is the position of the VT with the index $(l,j,k)$ in Cartesian coordinates. The number of VTs is not necessarily equal to the number of measurements due to the existence of spurious measurements (false alarms that are not coming from any VTs) and the absence of measurements (missed detections that should have come from VTs). The association between the measurement and VT $c_t^{l,j,k} = m$ means that the VT $\mathbf{v}^{l,j,k}$ is associated with the measurement $\mathbf{z}^{m,j,k}(t)$. 

The state vector of the vehicle can be represented as 
\begin{align}
\begin{split}
    \mathbf{r}_{1:t} &= \left[ \mathbf{r}_1,\ldots, \mathbf{r}_t\right]
\end{split}\\
\begin{split}
    \mathbf{r}_t  &= \left[r_x(t), r_y(t), r_z(t),r_{\psi}(t), r_{\theta}(t), r_{\phi}(t)\right]^T
    \end{split}
\end{align}
here $\mathbf{r}_p(t) = [r_x(t), r_y(t), r_z(t)]^T$ is the position of the vehicle in Cartesian coordinates at time index $t$, and $[r_{\psi}(t), r_{\theta}(t),r_{\phi}(t)]$ are the yaw, pitch, and roll of the vehicle.    

The vehicle's velocity $\mathbf{u}_t$ includes longitudinal, lateral, vertical, yaw, pitch, and roll velocities. Rotational velocities are observed by the inertial measurement unit (IMU), and longitudinal speed can also be observed with wheel odometry. The velocity can be represented as
\begin{equation}
    \mathbf{u}_t = \left[u_x,u_y,u_z,u_{\psi},u_{\theta}, u_{\phi}\right]^T.
\end{equation}

The FastSLAM algorithm in \cite{thrun2005pr}
is adopted to solve the posterior problem. If the data association is known (the method to acquire the data association is described at the end of \Cref{sec:fastslam_update}), FastSLAM can decompose the posterior into a factored form of 
\begin{align}\label{eq:general_eq}
\begin{split}
p\left(\mathbf{V},\mathbf{r}_{1:t}\mid\mathbf{Z}_{1:t},\mathbf{c}_{1:t},\mathbf{u}_{1:t} \right)= p\left(\mathbf{r}_{1:t}\mid\mathbf{Z}_{1:t},\mathbf{c}_{1:t},\mathbf{u}_{1:t}\right)\\
    \prod_{n \in\{K,J,L\}} p\left(\mathbf{v}_n\mid \mathbf{r}_{1:t},\mathbf{Z}_{1:t},\mathbf{c}_{1:t}\right).
\end{split}
\end{align}

Since MPCs from different antenna ports and different BSs can be separated by cell ID, they are independent and should not be associated, so the FastSLAM model is extended here and the posterior can be further factored as
\begin{align}
\begin{split}
&p\left(\mathbf{V},\mathbf{r}_{1:t}\mid\mathbf{Z}_{1:t},\mathbf{u}_{1:t},\mathbf{c}_{1:t}\right) \\
&= p\left(\mathbf{r}_{1:t}\mid\{\mathbf{Z}_{1:t}^{j,k},\mathbf{c}_{1:t}^{j,k}\}_{j\in J, k\in K},\mathbf{u}_{1:t}\right)\\    
&\prod_{k=1}^K\prod_{j=1}^J\prod_{l=1}^{L} p\left(\mathbf{v}^{l,j,k}\mid \mathbf{r}_{1:t},\mathbf{Z}_{1:t}^{j,k},\mathbf{c}_{1:t}^{l,j,k}\right)
\end{split}
\end{align}
here $\mathbf{v}^{l,j,k}$ represents the $l$-th VT from the $j$-th antenna port of the $k$-th BS, and $\mathbf{Z}_{1:t}^{j,k}$ and $\mathbf{c}_{1:t}^{j,k}$ represent the measurements and association of the MPCs from the $j$-th antenna port of the $k$-th BS. $\{\mathbf{Z}_{1:t}^{j,k},\mathbf{c}_{1:t}^{j,k}\}_{j\in J, k\in K}$ represents the combination of the measurements and the correspondence that is constrained to the MPCs from the same antenna port and BS. This method can process the position estimation of VTs from different antenna ports and BSs separately. It reduces computational complexity and provides flexibility to add or remove BSs. 
\section{Extended FastSLAM update with the estimated MPC parameters}\label{sec:fastslam_update}
Vehicle pose evolves as a function of control inputs and physical motion constraints, and it is defined as the motion model 
\begin{align}
    p(\mathbf{r}_t\mid\mathbf{r}_{t-1},\mathbf{u}_t)
\end{align}
here $\mathbf{r}_t$ is a probabilistic function of the vehicle's control input $\mathbf{u}_t$ and the previous pose state $\mathbf{r}_{t-1}$. 

The FastSLAM algorithm employs a particle filter \cite{Arulampalam2002tsp} to estimate the vehicle pose posterior. At each time index, it preserves a set of particles representing the posterior $p(\mathbf{r}_{1:t}\mid\{\mathbf{Z}_{1:t}^{j,k},\mathbf{c}_{1:t}^{j,k}\}_{j\in J, k\in K},\mathbf{u}_{1:t})$, and the set is denoted as $\mathbf{R}_{1:t}$. Each particle $\mathbf{r}_{i,1:t}$ represents the $i$-th hypothesis of the vehicle's path, i.e.,
\begin{equation}
    \mathbf{R}_{1:t} = \{ \mathbf{r}_{i,1:t} \}_i = \{ \mathbf{r}_{i,1},\ldots, \mathbf{r}_{i,t}\}_i.
\end{equation}

The particle $\mathbf{r}_{i,t-1}$ at time index $t-1$ is used to generate a probabilistic hypothesis of the vehicle's pose $\mathbf{r}_{i,t}$ at time index $t$ by sampling from the probabilistic motion model
\begin{equation}
    \mathbf{r}_{i,t} \sim p\left(\mathbf{r}_{t}\mid\mathbf{r}_{i,t-1},\mathbf{u}_t\right).
\end{equation}

After each particle is generated, the FastSLAM algorithm updates the posterior over the VT estimates associated with each particle. For the VT connected to the $i$-th particle of vehicle state, if there is no clearly associated observation, then it will keep the status unchanged, otherwise, the posterior at the time index $t$ will be updated as follows
\begin{align}
\begin{split}
    &p\left(\mathbf{v}_i^{l,j,k} \mid \mathbf{r}_{i,1:t},\mathbf{Z}_{1:t}^{j,k},\mathbf{c}_{i,1:t}^{l,j,k}\right)= \\
    & \eta p\left(\mathbf{z}_{t}^{c_{i,t}^{l,j,k},j,k} \mid \mathbf{r}_{i,t},\mathbf{v}_{i}^{l,j,k},{c}_{i,t}^{l,j,k}\right) \\
    &p\left(\mathbf{v}_i^{l,j,k} \mid \mathbf{r}_{i,1:t-1},\mathbf{Z}_{1:t-1}^{j,k},\mathbf{c}_{i,1:t-1}^{l,j,k}\right) 
\end{split} 
\end{align}
here $\eta$ is the normalization factor, and the posterior of $\mathbf{v}_i^{l,j,k}$ at the moment $t-1$ is assumed to be Gaussian with the following mean and variance. 
\begin{equation} \label{eq:vt_pos}
\resizebox{0.88\hsize}{!}{%
    $p\left(\mathbf{v}_i^{l,j,k} \mid \mathbf{r}_{i,1:t-1},\mathbf{Z}_{1:t-1}^{j,k},\mathbf{c}_{i,1:t-1}^{l,j,k}\right) \sim \mathcal{N}\left(\boldsymbol{v}_i^{l,j,k};\boldsymbol{\mu}_{i,t-1}^{l,j,k},\boldsymbol{\Sigma}_{i,t-1}^{l,j,k}\right).$%
    }
\end{equation}
To ensure that the estimate of VT at the time index $t$ is Gaussian, FastSLAM linearizes the perceptual model $p(\mathbf{z}_{t}^{c_{i,t}^{l,j,k},j,k} \mid \mathbf{r}_{i,t},\mathbf{v}_{i,t}^{l,j,k},{c}_{i,t}^{l,j,k})$, and the measurement function can be approximated by Taylor expansion as
\begin{align}
\begin{split} \label{eq:lineriz}    h\left(\mathbf{v}_i^{l,j,k},\mathbf{r}_{i,t}\right) &= \hat{\mathbf{z}}_{i,t}^{l,j,k} + \mathbf{H}_{i,t}^{l,j,k}\left(\mathbf{v}_{i}^{l,j,k} - \boldsymbol{\mu}_{i,t-1}^{l,j,k}\right)
\end{split}\\
\begin{split}     
     \hat{\mathbf{z}}_{i,t}^{l,j,k} &= h\left(\boldsymbol{\mu}_{i,t-1}^{l,j,k},\mathbf{r}_{i,t}\right) 
\end{split}   
\end{align}
here the function $h$ is defined to estimate the distance, azimuth AOA, and elevation AOA from the positions of the vehicle and the VT, and $\mathbf{H}_{i,t}^{l,j,k}$ is the Jacobian of $h$. The function $h$ is defined as follows
\begin{align}  
\begin{split}
    \hat{d}_i^{l,j,k}(t) &= \left\Vert \boldsymbol{\mu}_{i,t-1}^{l,j,k}-\mathbf{r}_{i,p}(t) \right\Vert          
     \end{split}\\
     \begin{split}
     \hat{\varphi}_i^{l,j,k}(t) &=  \text{atan}\left(\frac{\hat{y}}{\hat{x}}\right)    
     \end{split}\\
     \begin{split}
     \hat{\theta}_i^{l,j,k}(t) &=  \text{asin}\left(\frac{\sqrt{\hat{x}^2+\hat{y}^2}}{\hat{z}}\right) 
     \end{split}
\end{align}
here $\left[\hat{x}, \hat{y}, \hat{z}\right]^T$ is acquired by applying Euler's rotation theorem \cite{eulermatrix} with the rotation matrix $\mathbf{R}\left(r_{i,\psi}(t), r_{i,\theta}(t), r_{i,\phi}(t)\right)$ 
{\small
\begin{align}
    \left[\hat{x}, \hat{y}, \hat{z}\right]^T = \mathbf{R}\left(r_{i,\psi}(t), r_{i,\theta}(t), r_{i,\phi}(t)\right) (\boldsymbol{\mu}_{i,t-1}^{l,j,k}-\mathbf{r}_{i,p}(t)).
\end{align}
}%
With the approximation, the mean and covariance of the VT at time index $t$ can be updated with the standard EKF \cite{Kay97est_theory} as follows
{\small
\begin{align}
    \begin{split}
        \mathbf{K}_{i,t}^{l,j,k} &= \boldsymbol{\Sigma}_{i,t-1}^{l,j,k}{\mathbf{H}_{i,t}^{l,j,k}}^T\left(\mathbf{H}_{i,t}^{l,j,k}\boldsymbol{\Sigma}_{i,t-1}^{l,j,k}{\mathbf{H}_{i,t}^{l,j,k}}^T+\mathbf{Q}_t\right)^{-1}
    \end{split}\\
    \begin{split}
    \boldsymbol{\mu}_{i,t}^{l,j,k} &= \boldsymbol{\mu}_{i,t-1}^{l,j,k} +\mathbf{K}_{i,t}^{l,j,k}(\mathbf{z}_{t}^{c_{i,t}^{l,j,k},j,k}-\hat{\mathbf{z}}_{i,t}^{l,j,k})
    \end{split}\\
    \begin{split}        
    \boldsymbol{\Sigma}_{i,t}^{l,j,k} &= (\mathbf{I}-\mathbf{K}_{i,t}^{l,j,k}\mathbf{H}_{i,t}^{l,j,k})\boldsymbol{\Sigma}_{i,t-1}^{l,j,k}.
    \end{split}
\end{align}
}%
After the posterior of the VTs is updated, the importance factors of all the particles are calculated and used to resample the particles proportionally. The calculation of the importance factor of the $i$-th particle is given as follows
\begin{align}
\begin{split}
    w_{i,t} =& \frac{\textrm{target distribution}}{\textrm{proposal distribution}}\\
    =& \frac{p(\mathbf{r}_{i,1:t}\mid\{\mathbf{Z}_{1:t}^{j,k},\mathbf{c}_{i,1:t}^{j,k}\}_{j\in J, k\in K},\mathbf{u}_{1:t})}{p(\mathbf{r}_{i,1:t}\mid\{\mathbf{Z}_{1:t-1}^{j,k},\mathbf{c}_{i,1:t-1}^{j,k}\}_{j\in J, k\in K},\mathbf{u}_{1:t})}\\
    \propto &\prod_{k\in K} \prod_{j\in J} \prod_{l\in L} \int p\left(\mathbf{z}_t^{c_{i,t}^{l,j,k},j,k} \mid \mathbf{r}_{i,t}, \mathbf{v}_{i}^{l,j,k}, {c}_{i,t}^{l,j,k} \right) \\    &p\left(\mathbf{v}_{i}^{l,j,k}\mid\mathbf{r}_{i,1:t-1},\mathbf{Z}_{1:t-1}^{j,k},\mathbf{c}_{i,1:t-1}^{l,j,k}\right)d\mathbf{v}_{i}^{l,j,k}.
\end{split}
\end{align}
The last part in the equation is already defined in \cref{eq:vt_pos}. With the same linearization as in \cref{eq:lineriz}, the importance factor can be calculated as
\begin{align}
\begin{split}
 &w_{i,t} \approx \eta \prod_{k\in K} \prod_{j\in J} \prod_{l\in L}\left| 2 \pi \mathbf{Q}_{i,t}^{l,j,k}\right| ^{-\frac{1}{2}} \\
 &e^{-\frac{1}{2}(\mathbf{z}_{t}^{c_{i,t}^{l,j,k},j,k} - \hat{\mathbf{z}}_{i,t}^{l,j,k})^T (\mathbf{Q}_{i,t}^{l,j,k})^{-1}(\mathbf{z}_t^{c_{t}^{l,j,k},j,k} - \hat{\mathbf{z}}_{i,t}^{l,j,k})} 
 \end{split}
\end{align}
and the covariance is 
\begin{equation}
    \mathbf{Q}_{i,t}^{l,j,k} = \left( \mathbf{H}_{i,t}^{l,j,k} \right)^T \boldsymbol{\Sigma}_{i,t-1}^{l,j,k} \mathbf{H}_{i,t}^{l,j,k} + \mathbf{Q}_t
\end{equation}
here $\mathbf{Q}_t$ is the covariance matrix of the measurement. Since the SNR is related to the accuracy of the estimated parameters, it is used to update the covariance matrix of the measurement, and $\mathbf{Q}_t$ can be written as 
\begin{equation}
    \mathbf{Q}_t = \mathbf{Q} \cdot \textrm{diag}(\boldsymbol{\lambda}_t)
\end{equation}
here $\textrm{diag}(\boldsymbol{\lambda}_t)$ is the diagonal matrix constituted of the elements of the SNR vector $\boldsymbol{\lambda}_t$.

In the FastSLAM with maximum likelihood data association, the association ${c}_{i,t}^{l,j,k}$ is determined by maximizing the following likelihood 
{\small
\begin{align}
\begin{split}
    &{c}_{i,t}^{l,j,k}= \argmax_{l'} p\left(z_t^{l',j,k} \mid l',\mathbf{c}_{i,1:t-1}^{l,j,k},\mathbf{r}_{i,1:t},\mathbf{Z}_{1:t-1},\mathbf{u}_{1:t}\right)\\   &=\argmin_{l'}\left((\mathbf{z}_{t}^{l',j,k} - \hat{\mathbf{z}}_{i,t}^{l,j,k})^T (\mathbf{Q}_{i,t}^{l,j,k})^{-1}(\mathbf{z}_t^{l',j,k} - \hat{\mathbf{z}}_{i,t}^{l,j,k})\right).
 \end{split}
\end{align}
}%
For multiple VTs and multiple measurements, the Hungarian algorithm \cite{Kuhn1955Hungarian} is applied to find the maximum likelihood data association among them.
\section{measurement setup and SLAM results analysis}\label{sec:measure}
\begin{figure}	\centerline{\includegraphics[scale=0.105]{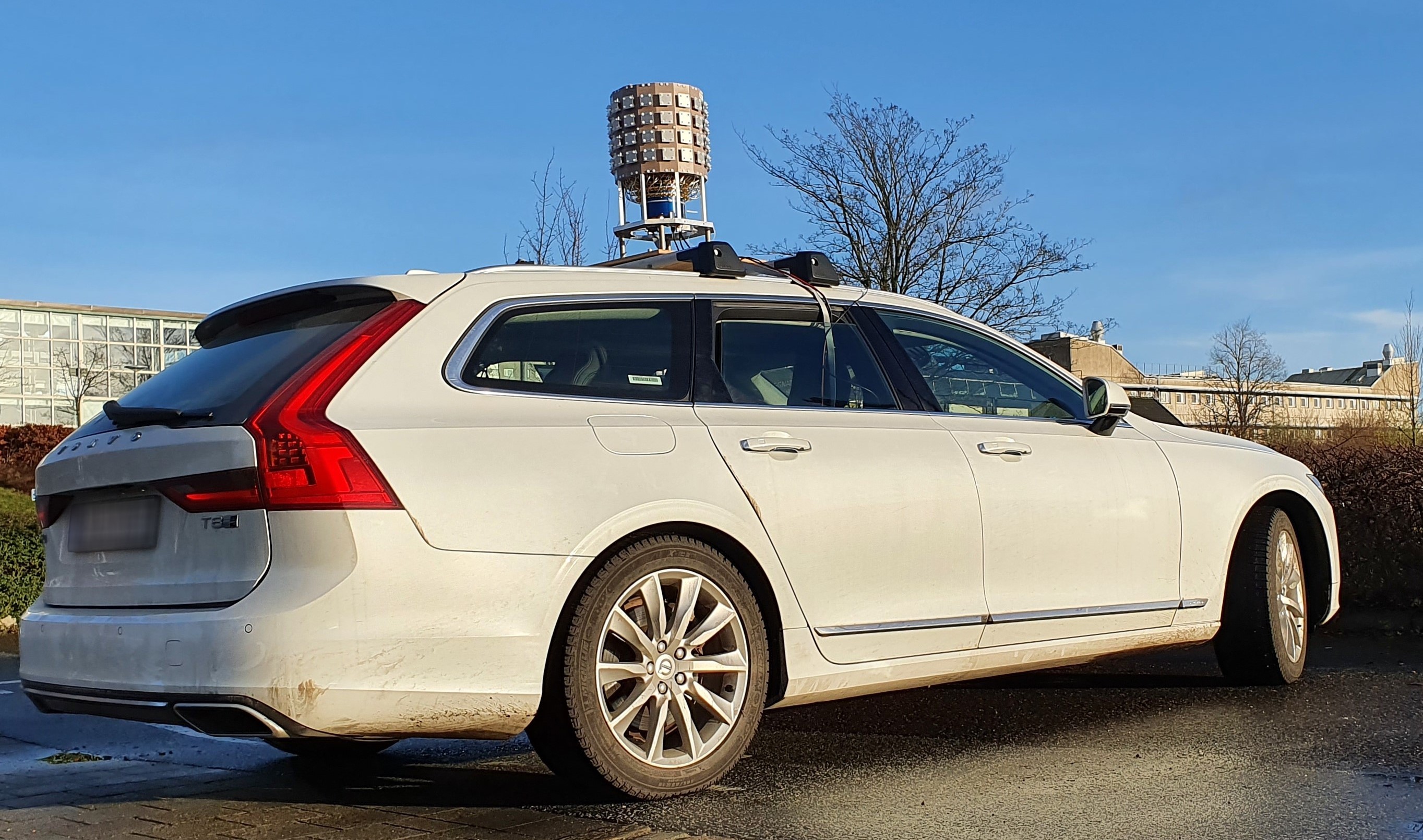}}
	\caption{The massive antenna array on top of the measurement vehicle \cite{junshi2022ic_rimax}.}
	\label{fig:volvo_car}
\end{figure}

A measurement system with a USRP controlling the 128-port stacked uniform circular antenna array mounted on the roof of a vehicle is shown in \Cref{fig:volvo_car}. The system was used to receive and log CRS symbols from commercial LTE BSs in the city of Lund, Sweden. A rubidium standard disciplined by GPS beforehand was used as a stable frequency reference for the USRP to minimize clock drift, and the clock offsets between different BSs and the vehicle were assumed to be unknown constants. An OXTS RT3003G \cite{oxts} was used for ground truth position and orientation of the vehicle and the antenna array. The GPS receiver inside the USRP was used for time alignment between ground truth and data logging. Yaw velocity observations from the IMU and the longitudinal speed observations from wheel odometry were used as the input velocity of the extended FastSLAM algorithm, and the vertical, pitch, and roll velocities were assumed to be zero owing to the flat terrain and constrained vehicle dynamics. The parameters for the measurement system are listed in \Cref{tab:measure_setup}.

\begin{table}
  \begin{center}
    \caption{Measurement system information}\label{tab:measure_setup}
        \begin{tabular}{|l|l|}
            \hline
            \textbf{Parameter Name} & \textbf{Value} \\  
            \hline
            Center frequency & 2.66 GHz  \\
            \hline
            System bandwidth & 20 MHz  \\
            \hline
            BS number & 2 \\            
            \hline
            Cell IDs of BS A & 375, 376, 377\\
            \hline
            Cell IDs of BS B & 177, 178, 179\\              
            \hline
            Tx antenna port number & 2 \\
            \hline
            Rx antenna port number & 128  \\            
            \hline
            Snapshot interval & 75 ms \\
            \hline
            Total snapshot number & 6850 \\
            \hline
            Total test time & 8.5 minutes \\
            \hline
            Traversed distance & 530 meters \\     
            \hline            
    \end{tabular}
  \end{center}
  \vspace{-7mm}
\end{table}

\begin{figure*}
\centerline{\includegraphics[width=17cm]{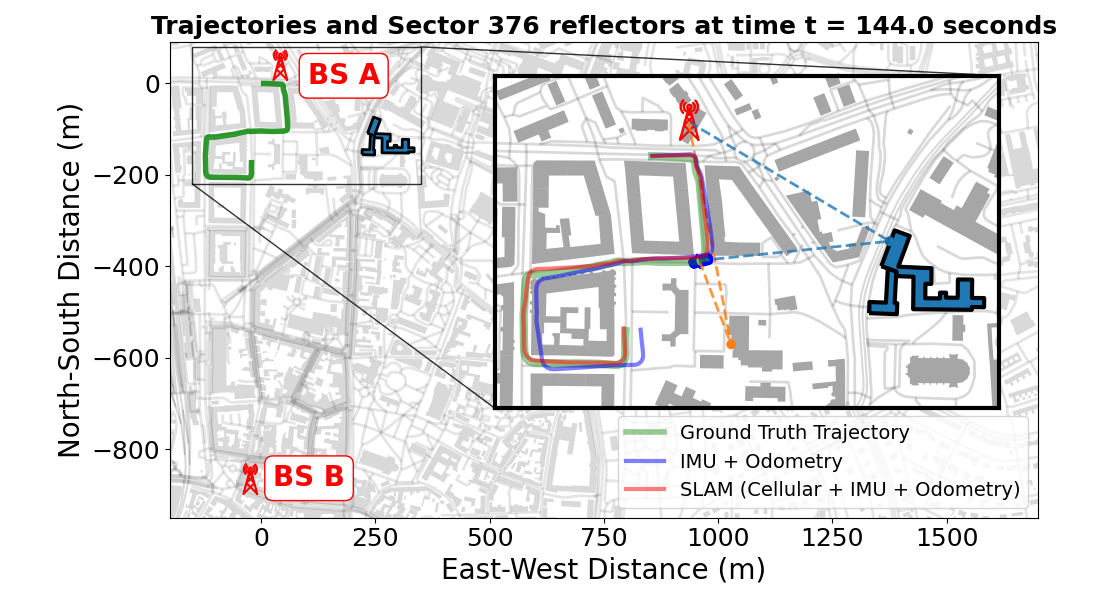}}
\caption{The ground truth trajectory together with the SLAM estimation and proprioception-only, and the positions of reflectors from one sector at one time index. The building that provides the long-lived VT is emphasized with a darker color and outline.}
\label{fig:slam_trj_vt}
%\vspace{-4mm}
\end{figure*}
\begin{figure*}
\centering
\begin{subfigure}{0.4\textwidth}
  \centerline{\includegraphics[scale=0.63]{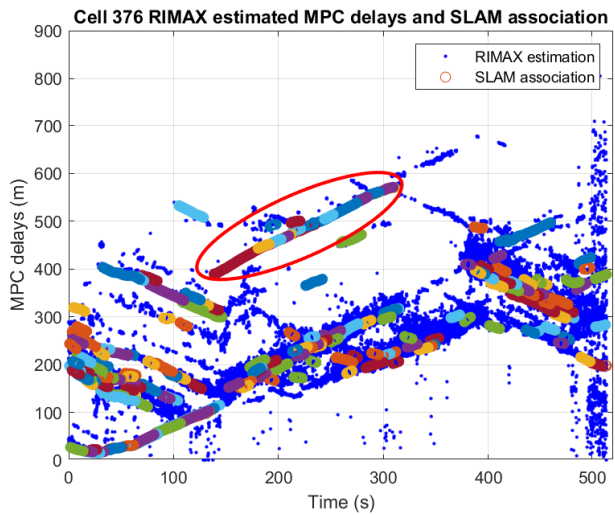}}
\subcaption{Cell 376 multipath component delays estimated by RIMAX and associated with SLAM.}
\label{fig:da_376}
\end{subfigure} 
\hspace{15mm}
\begin{subfigure}{0.4\textwidth}
  \centerline{\includegraphics[scale=0.63]{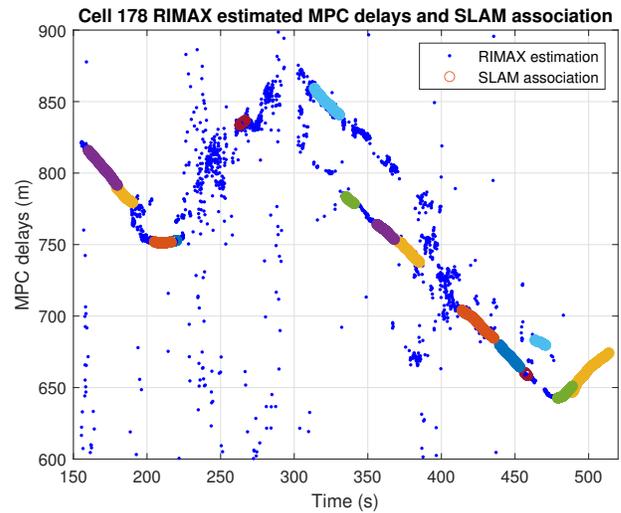}}
\subcaption{Cell 178 multipath component delays estimated by RIMAX and associated with SLAM.}
\label{fig:da_178}
\end{subfigure}
\caption{Multipath component delays estimated by RIMAX and associated with SLAM for sectors 376 and 178}
\end{figure*}

The measurement trajectory is shown in \Cref{fig:slam_trj_vt}. The outer figure gives an overview of the relative positions of the BSs and trajectory, and the inset plot gives a more detailed view of the trajectories of the ground truth, SLAM estimation fusing cellular signals with IMU and wheel odometry, and proprioception using the IMU and wheel odometry alone. The estimated positions of the virtual transmitters from SLAM are mapped to physical reflectors with an assumption of first-order reflection, and the physical reflectors are also shown in the figure as dots. A particularly noteworthy long-lived NLOS MPC is shown inside the red ellipse in \Cref{fig:da_376}. This MPC is mapped to the physical environment as a blue dot shown in \Cref{fig:slam_trj_vt} and the associated building is plotted with darker colors. The reflections come from a wall 230 meters away from the BS, and they are 170-350 meters away from the vehicle as it drives apart. It shows the potential of using NLOS MPCs for positioning in complicated urban environments

\begin{figure}[h]
 \vspace{-2mm}
\centerline{\includegraphics[scale=0.5]{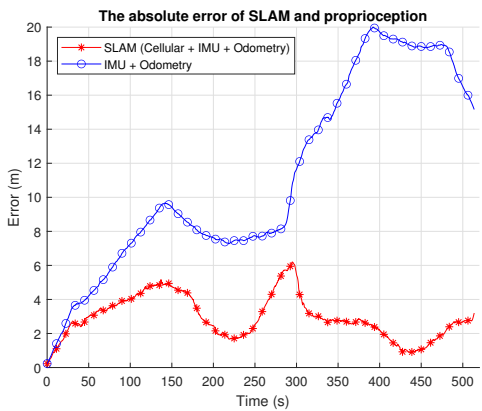}}
\caption{The absolute error of SLAM and proprioception.}
\label{fig:abs_error}
\vspace{-5mm}
\end{figure}

The absolute error of the estimated vehicle trajectory from SLAM and proprioception only are shown as a function of time in \Cref{fig:abs_error}. It can be observed that the extended FastSLAM can greatly improve positioning performance. It has a maximum absolute horizontal error of 6 meters after 290 seconds and has 3 meters of horizontal error after a total traversed distance of 530 meters, while the absolute error of the IMU and wheel odometry alone have a maximum horizontal error of 20 meters and 15 meters at the end of the measurement. 

The MPC delay estimates from RIMAX for sector 376 of BS A and sector 178 of BS B are shown in \Cref{fig:da_376} and \Cref{fig:da_178} respectively. The associated MPC delays from one particle of the SLAM are also shown in the corresponding figures. It can be observed that the extended FastSLAM can associate the estimated MPC delays accurately, while effectively suppressing spurious measurements.
\begin{comment}
\begin{figure}
\centerline{\includegraphics[scale=0.63]{image/Cell_376_rimax_vs_slam_association.eps}}
\caption{Cell 376 multipath component delays estimated by RIMAX and associated with SLAM.}
\label{fig:da_376}
%\vspace{-5mm}
\end{figure}
\begin{figure}
\centerline{\includegraphics[scale=0.60]{image/Cell_178_rimax_vs_slam_association.eps}}
\caption{Cell 178 multipath component delays estimated by RIMAX and associated with SLAM.}
\label{fig:da_178}
\vspace{-4mm}
\end{figure}
\end{comment}
%\begin{comment}
%\end{comment}
\section{conclusion}\label{sec:summary}
In this paper, an extended FastSLAM algorithm using multipath component delays and angular information is developed, which simplifies the data association problem and processes the multipath components from different antenna ports and base stations independently. The multipath component delays and angular information extracted from the commercial LTE signals received by the 128-port antenna array are processed by the extended FastSLAM algorithm, and the results validate the algorithm and demonstrate the capability of using cellular signals for high accuracy positioning in complicated urban environments. 
\section*{acknowledgement}
This work was financed in part by the Swedish Innovation Agency
VINNOVA through the MIMO-PAD Project (Reference number
2018-05000). Computational resources were provided by the Swedish National Infrastructure for Computing (SNIC) at HPC2N, partially funded by the Swedish Research Council through grant agreement no. 2018-05973.

\balance

%\bibliographystyle{ieeetr}
%\bibliography{MyCite}

\bibliographystyle{IEEEtran}
\bibliography{MyCite}

\end{document}